\documentclass[11,pdftex]{article}

\usepackage{tikz}
\usepackage{amsmath}
\usetikzlibrary{arrows}
\usetikzlibrary{snakes}
\usetikzlibrary{shapes}
\usepackage{subfigure}
\usepackage{hyperref}

\newcommand{\ie}{\textit{ i.e. }}
\title{Simulations of the spread of COVID-19 and control policies in Tunisia}

\author{Slimane BenMiled $^{1}$ and Amira Kebir $^{1,2}$ }

\date{%
$^{1}$ \quad BIMS Lab, Institut Pasteur de Tunis, Universit\'e de Tunis el Manar; slimane.benmiled@fst.utm.tn\\
$^{2}$ \quad IPEIT, Universit\'e de Tunis; amira.kebir@pasteur.utm.tn}

\begin{document}
\maketitle
\abstract{ We develop and analyze in this work an epidemiological model for COVID-19 using Tunisian data. Our aims are first to evaluate Tunisian control policies for COVID-19 and secondly to understand the effect of different screening, quarantine and containment strategies and the rule of the asymptomatic patients on the spread of the virus in the Tunisian population.  With this work, we show that Tunisian control policies are efficient in screening infected and asymptomatic individuals and that if containment and curfew are maintained the epidemic will be quickly contained. }

\section{Introduction}

On March 11, 2020, WHO announced that the COVID-19 epidemic had passed the pandemic stage, indicating its autonomous spread over several continents. Since March 22, Tunisia has experienced a turning point and general health containment has begun. Tunisia's strategy of containment and targeted screening 
corresponds to the first WHO guidelines, the aim being to detect clusters by diagnosing only suspicious persons and then to trace the people who came into contact with the positive cases. 
This method is now showing its limitations. The mass screening carried out in some countries shows that asymptomatic patients or those who develop only a mild form of the disease may exist in significant numbers.  So what is the rule of the asymptomatic patients on the spread of the virus in the Tunisian population and does containment and mass screening strategies are sufficient to control the spread of the virus in the Tunisian population?

In this work, a mathematical epidemiological model for COVID-19 is developed to study and predict the effect of different screening, quarantine, and containment strategies on the spread of the virus in the Tunisian population. This model is more detailed than the classical model (SIR) but it remains very simple in its structure. Indeed, all individuals are assumed to react on average in the same way to the infection, there are no differences in age, sex, contacts.  The model is calibrated and fitted to Tunisian data.

In what follows, we present the model and its assumptions. Then we calibrate different parameters of the model based on the Tunisian data and calculate the expression of the basic reproduction number $R_0$ as a function of the model parameters. Finally, we carry out simulations of interventions  and compare different strategies for suppressing and controlling the epidemic. 

\section{Model description}
\begin{figure}[tbp]
\scalebox{1.5}{\begin{tikzpicture}
		\node [draw,rectangle, color=blue] (0) at (-7.5, 2) {S};
		\node [draw,rectangle,color=red] (1) at (-5, 1) {Q};
		\node [draw,rectangle,color=red] (2) at (-7.5, -0) {As};
		\node [draw,rectangle,color=red] (3) at (-5, -1) {I};
		\node [draw,rectangle] (4) at (-2.5, 1) {D};
		\node [draw,rectangle, color=blue] (5) at (-2.5, -1) {R};
		\draw[->]  (0) -- node[midway,left]{$\alpha_1 As+\alpha_2 I$} (2) ;
		\draw [->] (2) -- node[midway, above]{$\tau_1$} (1);
		\draw [->] (2) -- node[midway,below]{$\beta$} (3);
		\draw [->] (1) -- node[near start,above]{$\gamma$} (5);
		\draw [->] (1) -- node[midway,above]{$\mu$} (4);
		\draw [->] (3) -- node[midway,left]{$\tau_2$} (1);
		\draw [->] (3) -- node[midway,below]{$\gamma$} (5);
		\draw [->] (3) -- node[near start,below]{$\mu$} (4);
\end{tikzpicture}}
 \caption{The model flow chart}
 \label{fig:1}
\end{figure}
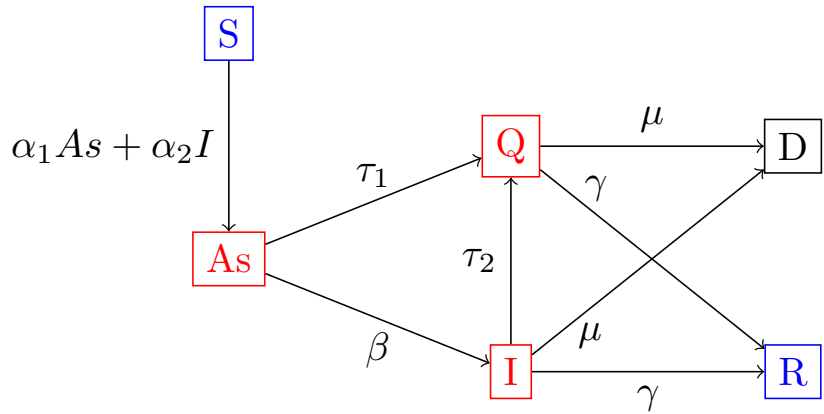

COVID-19 is a respiratory disease that spreads mainly through the respiratory droplets expelled by people who cough. So the transmission is usually direct from person to person.  Infection is considered possible even when in contact with a person with mild symptoms. In fact, in the early stages of the disease, many people with the disease have only mild symptoms.  It is, therefore, possible to contract COVID-19 through contact with a person who does not feel sick. Subsequently, in this work, we consider susceptible individuals, noted $S$, who are infected first go through a stage where they are infected but asymptomatic, noted $As$ for unreported asymptomatic infectious.  This stage appears to be particularly important in the spread of COVID-19.
 The individuals then develop symptoms and become symptomatic infectious, so either enter directly into a quarantine stage, noted $Q$, corresponds to reported symptomatic infectious individuals, or go through a moderate or severe infectious stage, noted $I$ for unreported symptomatic infectious and then can go through the quarantine stage or not.  Finally, the infection ends and the individuals are then immunized, denoted $R$ or dead, denoted $D$. This life cycle can be represented using the following flow chart (\ref{fig:1}) followed by table \ref{table:1} that lists the model parameters.

\begin{table}[h]
\caption{Table of parameters}
\label{table:1}
\centering
\scalebox{0.7}{\begin{tabular}{cccc}
\hline
\textbf{Notation} & \textbf{Description} &  \textbf{Value} &  \textbf{Reference}  \\
\hline
\hline $t_0$ & Initial epidemic time & $-3.172$ & fitted  \\
 $S_0$ & Number of susceptible at time $t_0$ & $11791748$ & Tunisian population \\
 $As_0$ & Number of asymptomatic at time $t_0$ & $0.317$ & fitted \\
 $I_0$ & Number of infected at time $t_0$ & $0.050$ & fitted \\
 $\alpha_1$ & Transmission rate by asymptomatic & $4.513 e-08$ & fitted  \\
 $\alpha_2=f \alpha_1$ & Transmission rate by infected &  &   \\
  $f$ & & $3$  & estimated \\ 
 $\beta$ & Rate at which asymptomatic infectious  & $1/6$ & estimated from data  \\
  &  become  symptomatic infectious &  &  \\
 $\tau_1$ & Rate at which asymptomatic enter in quarantine & $0.400$ & fitted  \\
 $\tau_2$ & Rate at which infected enter in quarantine & $0.778$ & fitted  \\
 $\gamma$ & Rate of recovery & $0.045$ & estimated from data  \\
 $\mu$ & Rate of mortality & $0.003$ & estimated from data  \\
\hline
\end{tabular}}
\end{table}

The quarantine is assumed to act on the $As$ and $I$ stages. Indeed, we assumed that the state can detect  asymptomatic individuals by for example random screening, and then positive testing ones go into quarantine. Let $\tau_1$, respectively $\tau_2$, be the quarantine rate for $As$ class, respectively $I$.  

We assumed that the  asymptomatic individual, $As$, turn out to be $I$ at a rate $\beta$. We further assumed that quarantined individuals, $Q$ and infected individuals, $I$ either die at a rate of $\mu$ per unit of time or become recovered/immune, $R$, at a rate of $\gamma$ per unit of time.  

Finally, we assume that each healthy individual is infected proportionally by $As$ asymptomatic individuals, with a rate of $\alpha_1$ and by $I$ infected individuals, with a rate of $\alpha_2$. As $I$ state is constituted with the moderate or severe state, they are more contagious than the $As$ state, therefore, $\alpha_1<\alpha_2$.

Therefore, our model consists of the following system of $6$ ordinary differential equations:

\begin{equation}
 \label{eq:model}
 \left\{
 \begin{array}{lcl}
 \displaystyle  \frac{dS}{dt}&=&-\alpha_1SAs-\alpha_2SI \\
 \displaystyle  \frac{dAs}{dt}&=&\alpha_2SI+(\alpha_1S-\beta-\tau_1)As \\
  \displaystyle \frac{dI}{dt}&=&\beta As-(\tau_2+\gamma+\mu)I\\
  \displaystyle  \frac{dQ}{dt}&=&\tau_1 As+\tau_2 I-(\mu+\gamma)Q\\
   \displaystyle  \frac{dR}{dt}&=&\gamma (I+Q)\\
  \displaystyle   \frac{dD}{dt}&=&\mu(I+Q) 
   \end{array}
\right.
\end{equation}

With an initial condition at time $t=t_0$ defined as following:

$$S(t_0)=S_0>0, As(t_0)=As_0>0, I(t_0)=I_0>0, Q(t_0)=0, R(t_0)=0, D(t_0)=0$$

One of the advantages of the basic reproduction number $R_0$ concept is that it can be calculated from the moment the life cycle of the infectious agent is known. We calculate the $R_0$ for our model  using the Next Generation Theorem \cite{Diekmann1990} (see section \ref{sectionR0}),  

\begin{equation}\label{R0}
R_0=\frac{\alpha_1 S_0}{\beta+\tau_1}+\frac{\alpha_2\beta S_0}{(\beta+\tau_1)(\tau_2+\gamma+\mu)}
\end{equation}


The first term of the expression of (\ref{R0}) corresponds to infections generated by asymptomatic types (healthy carrier to mild symptoms). The second term corresponds to secondary infections caused by moderate or severe symptomatic infection.
With this expression, we can see that there are several ways to lower the $R_0$ and thus control the epidemic. For example, we can reduce the number of susceptible people  (decreasing $\alpha_1$ and $\alpha_2$) by confining the population, reducing contacts, and wearing masks. We can also reduce the rate of contact with an infected person by increasing quarantine rates ($\tau_1$ and $\tau_2$) by isolating asymptomatic or symptomatic infected persons through mass screening.

\section{Comparison with data}
The estimation of the different parameters of the model is done in three steps (see section \ref{estimation}). In the first step, we will estimate the start date of the epidemic, $t_0$, the initial states $ As(t_0)$ and $I(t_0)$  as well as the infection rates $\alpha_1$ and $\alpha_2$. In the second step, we estimate the mortality rate, $\mu$, and the recovery rate, $\gamma$. In the third step, we evaluate the parameters $\tau_1$ and $\tau_2$ by an optimization method. The program is available for download\footnote{\url{https://github.com/MayaraLatrech/covid19_sasymodel.git}}.

We used the  Tunisian Health Commission \footnote{https://covid-19.tn/} data-set   of reported data to model the epidemic in Tunisia. It represents the daily new-cases, death,  and recoveries in Tunisia. The first case was detected on March 2, 2020. 

To estimate the initial conditions $As(t_0)$ and $I(t_0)$ and  parameters $\alpha_1$ and  $\alpha_2$, we fix $S_0 = 11694720$, which corresponds to the total  population of Tunisia and assume that the variation in $S(t)$ is small during the period consider. We also fix the parameters $\beta, \gamma, \tau_1$ and $\tau_2 $. 
For this estimation, we adapt the method developed by \cite{Liu2020} in our case.  Let  $CRt)$ the cumulative number of reported  infectious cases at time $t$, defined by, 
\begin{equation}
 \label{eq:2}
CR(t)=\int_{t_0}^t\tau_1 As(t)+\tau_2 I(t) dt
\end{equation}

Let's assume that $CR(t)=\chi_1 \exp(\chi_2 t)-\chi_3$ with $\chi, \chi_2$ and $\chi_3$ three positive parameters that we estimate using log-linear regression on cases data (see figure \ref{fig:2} and table \ref{tab:2}).

\begingroup
\centering
\begin{figure}[htbp]
\centering
\subfigure[Fitted $CR$.]{\includegraphics[width=0.4\textwidth]{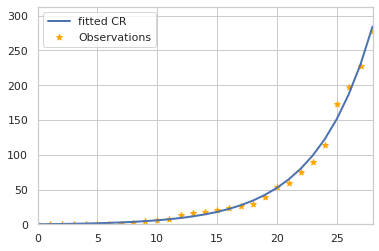}}\label{fig:2a}
\subfigure[Fitted $log(CR)$.] {\includegraphics[width=0.4\textwidth]{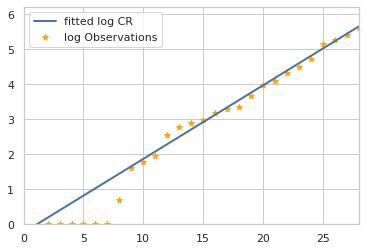}}\label{fig:2b}
\caption{The fitted  cumulative number of reported  infectious cases $CR= \chi_1 \exp(\chi_2 t)-\chi_3$ to Tunisian data using Linear regression.} \label{fig:2}
\end{figure}
\endgroup

\begin{table}[h]
\caption{Parameters of the estimated  cumulative number of reported  infectious cases $CR$}
\label{tab:2}
\centering
\begin{tabular}{ccccc}
\hline
$\chi_1$& $\chi_2$&$\chi_3$&$t_0$ & $R2$\\
\hline$1.208$&$0.210$&$0.729$&$-2.401$ &$0.993$\\
\hline
\end{tabular}
\end{table}

We obtain the model starting time of the epidemic $t_0$ by assuming that $CR(t_0)=0 $ and therefore equation \eqref{eq:2} implies that:
\begin{equation}
 \label{eq:3}
 t_0=\frac{1}{\chi_2}(\ln(\chi_3)-\ln(\chi_1)).
\end{equation}

For now, we assume that $\alpha_2=f \alpha_1$ with $f$ a fixed parameter bigger than $1$ and let's $\tau=\tau_2/\tau_1$. Then, by following using the approach described in the step 1 of section \ref{estimation}, we have:
\begin{align}
I(t_0)&=\frac{\beta }{\chi_2+\tau_2+\beta \tau+\gamma +\mu}\frac{\chi_2 \chi_3}{\tau_1}\\                                   
As(t_0)&=  (1-\frac{\beta \tau}{\chi_2+\tau_2+\beta \tau+\gamma + \mu})\frac{\chi_2 \chi_3}{\tau_1}\\
 \alpha_1&=\frac{(\chi_2+\beta+\tau_1)}{(\frac{f\beta}{\chi_2+\tau_2+\gamma +\mu}+1)S_0} \\
 R_0&=\frac{(\chi_2+\beta+\tau_1)}{(\beta+\tau_1)}\frac{(1+\frac{f\beta }{\tau_2+\gamma+\mu})}{(1+\frac{f\beta}{\chi_2+\tau_2+\gamma +\mu})}
 \end{align}
 
 \begin{figure}
\centering
 \includegraphics[width=0.4\textwidth]{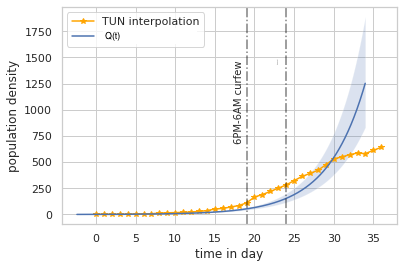}
   \caption{In this figure, we plotted  the number of individuals in quarantine without curfew  (in blue), $Q(t)$ (\ie the number of diagnosed cases per day), and the Tunisians data (in orange). }
   \label{fig:3}
\end{figure}

In figure \ref{fig:3}, we plotted the number of individuals in quarantine without curfew predicted by the ODE model, $Q(t)$ \ie the number of diagnosed cases per day and compared to the Tunisians data. We observe that from the 20th day onwards, the simulated curve deviates from the observed data. This deviation is due to the epidemic control policies put in place between 20 and 25 March (closure of cafes shops and the introduction of a curfew).

It can be seen that if the curfew had not been installed, the country could have had thousands of additional cases during the month of April.


\section{Numerical simulations}
\subsection{Effect of containment and curfew}
In figure \ref{fig:4} we study the effect of the curfew installed by the Tunisian state since March 20, 2020, on the number of infected people reported by the state.

Figure \ref{fig:4} shows the effect of two curfew strategies on the dynamics of the epidemic: a 12-hour curfew (the chosen Tunisian policy) and an 18-hour curfew (a more restrictive policy). During the period of curfew, the rate of infection $\alpha_1$ is divided by $100$.  In Figure \ref{fig:4a}, it can be seen that, for the chosen policy to maintains a 12-hour curfew for $100$ days, the epidemiological peak in terms of the number of reported infected is reached after about $50$ days with a value equal to $953$. After the peak, we observe a slow decrease in the number of reported infected persons. On the other hand, in the more restrictive case of$18$ hours curfew, the peak would be reached more quickly after $27$ days with a more rapid decrease. These values should be compared with the $774$ cases given by \cite{Abdeljaoued2020} and the fact that the epidemiological peak was reached around  April 29, 2020,  after about $58$ days with a reported infected number equal to $975$.

We represent on figure \ref{fig:4b}, the number of deaths by time, it appeared that the peak of the deaths is shifted to the peak of the infected for about $50$ days, this shift corresponds to the hypothesis that we made on the duration between the beginning of the symptoms and the deaths (\ie $30$ days).  We note that the simulated death curve overestimates the observed curve. This may be due to the mortality of unreported infected and therefore the surplus corresponds to unreported COVID-19 mortality, which makes the optimization of the model's variables for deaths imprecise. 

Moreover, the model predicts $228$ deaths at the end of the epidemic in the current case (12 hours curfew). In the case where the curfew was 18 hours, the number of deaths would be $84$. This information should be taken with caution, because at the time the simulations were made the number of deaths was low.

\begingroup
\centering
\begin{figure}[htbp]
\centering
\subfigure[Effect of curfew on number of reported  infected.]{\includegraphics[width=0.4\textwidth]{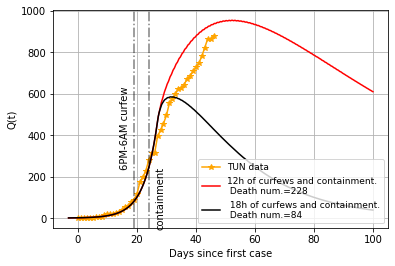}\label{fig:4a}}
\subfigure[Effect of curfew on number of death.] {\includegraphics[width=0.4\textwidth]{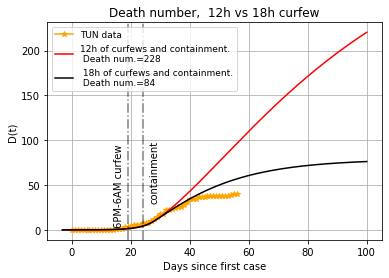}\label{fig:4b}}
\caption{Effect of containment on number of reported infected and death} \label{fig:4}
\end{figure}
\endgroup

\begin{figure}
\centerline{
 \includegraphics[width=0.5\textwidth]{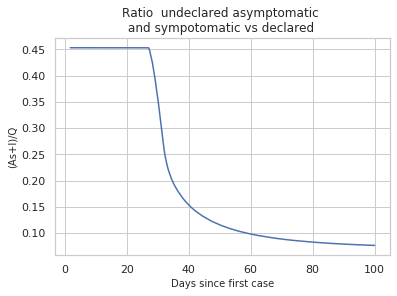}}
 \caption{Ratio of undeclared cases (asymptomatic and symptomatic)} \label{fig:5.5}
\end{figure}

 Finally, we notice that the ratio between the undeclared cases (asymptomatic and symptomatic) represents between $45 \%$ at the beginning of the epidemic for less than $10\%$ at the end of the epidemic (see figure \ref{fig:5.5}).

\subsection{Effect of mass screening}

\begin{figure}
\centerline{
  \includegraphics[width=0.5\textwidth]{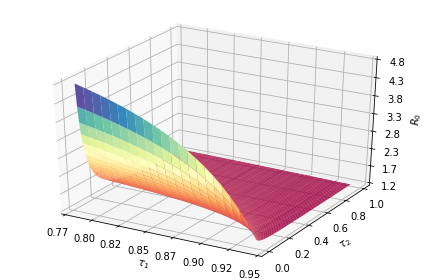}}
   \caption{Effect of quarantine rates, $\tau_1$ and $\tau_2$, on the basic reproduction number $R_0$. }\label{fig:6}
\end{figure}

\begin{figure}[htbp]
\centering
\subfigure[Effect of mass screening on number of reported  infected.]{\includegraphics[width=0.4\textwidth]{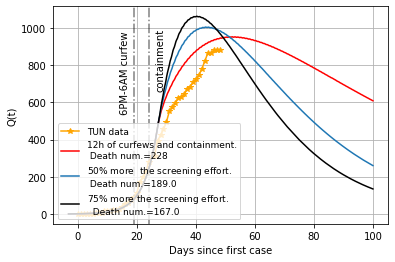}\label{fig:7a}}
\subfigure[Effect of mass screening on number of death.] {\includegraphics[width=0.4\textwidth]{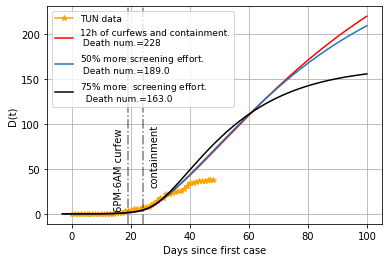}\label{fig:7b}}
\caption{Effect of different scenarios of mass screening policies on the tunisian population} \label{fig:7}
\end{figure}

We study in this section the effect of more intensive mass screening, \ie by increasing $\tau_1$ and $\tau_2$,  on the basic reproduction number, $R_0$, and on the number of declared infected, $Q$ (see figures \ref{fig:6}, \ref{fig:7}). 

In figure \ref{fig:6}, we can see that no matter how intensively we screen, we have $R_0>1$. Moreover, we note that to minimize the basic reproduction number, it is necessary to increase the massive screening of the class of asymptomatic infections. Similarly in figure \ref{fig:7}, we can see that the increase in the mass screening effort allows a more rapid decrease in the number of cases (see figure \ref{fig:7a}). This is probably since cases are detected earlier, they do not contribute to the contamination of the healthy ones. 

 Indeed, mass screening has an indirect effect on recruitment in the infect compartment. More specifically, we assume that $\beta+\tau_1$ does not vary when $\tau_1$ changes. Consequently, the effort of mass screening on asymptomatic patients cannot exceed this value, and then any additional effort beyond $\beta+\tau_1$ will be passed on to healthy patients and therefore will be useless.
 
  It is observed that the calibration of the model on the Tunisian data  using Metropolis-Hastings (MH) algorithm, gives a value of $\tau_1=0.78$, which represents a very important screening effort. This would prove that the screening strategy is very efficient. However, we didn't had access to the testing campaign methodology that would have allowed us to adjust our estimates.

Moreover, it is observed that the number of deaths at the end of the epidemic varies from $228$ to $167$,\ie a $20\%$ decrease in the number of cases (see figure \ref{fig:7b}).

\section{Conclusion}

It can be noted that since  April, 15,  2020, Tunisia has succeeded in slowing down the speed of propagation thanks to and containment and a curfew. With this work, we suggest that if containment and curfew are maintained, short-term projections could be more optimistic. The fact that the epidemic is quickly contained tends to show that the number of undeclared infected is low, which may suggest that our model is efficient for the evaluation of undeclared cases. In fact, we show that at the time of the epidemiological peak, the number of unreported infected persons constitutes at most $1/3$ of the infected population. However, this will need to be confirmed by a field evaluation.
Moreover, using Tunisian data, the optimization algorithm fixes the rate at which asymptomatic enter in quarantine, $\tau_1$, at a high value.  This expresses the good performance of the control policy of the Tunisian government. Indeed, in Tunisia, the control policy consists of an intense isolation campaign targeting sick individuals and their relatives. An effort of testing was carried out in a targeted manner, similar to snowball sampling. This approach enabled to have a major screening effort on infected and asymptomatic individuals. Finally,  the model was successfully able to predict the time of the peak at the end of April.

\appendix

\section{Materials and Methods}

\subsection{Computation of the basic reproduction number $R_0$}\label{sectionR0}
We use the next generation matrix to drive the basic reproduction number $R_0$ \cite{Diekmann1990}. In the system \eqref{eq:model} we have two infected compartments represented by the second and third equations of the system. Therefore, at the infection-free steady state, \ie for a small $(As, I)$ and $S=S_0$, the linear epidemic subsystem is :

\begin{equation}
 \label{eq:modellin}
 \left\{
 \begin{array}{lcl}
 \displaystyle \frac{dAs}{dt}&=&S_0(\alpha_1As+\alpha_2 I)-(\beta+\tau_1)As\\
 \displaystyle \frac{dI}{dt}&=&\beta As-(\tau_2+\gamma+\mu)I
   \end{array}
\right.
\end{equation}

If we set $X=(As, I)^T$ as the vector of infected, $F$ is the  matrix that represents the production of new infections and $T$ the matrix  of transfer into and out of the compartment by transmission, mortality, quarantine, and recovery, then the matrix form of the  linear epidemic subsystem is:

$$\dot{X}= (F-T)X$$

Where :
$F=\left(\begin{array}{cc}
\alpha_1 S_0& \alpha_2 S_0 \\ 
0 & 0
\end{array} \right)$
and  $T=\left(\begin{array}{cc}
\beta+\tau_1& 0 \\ 
-\beta & \tau_2+\gamma+\mu
\end{array} \right)$.

Therefore, the next generation matrix is :

$$FT^{-1}=\displaystyle\frac{S_0}{(\beta+\tau_1)(\tau_2+\gamma+\mu)}
\left(\begin{array}{cc}
 \alpha_1(\tau_2+\gamma+\mu)+\alpha_2\beta& \alpha_2 (\beta+\tau_1) \\ 
0 & 0
\end{array} 
\right)
$$

Knowing that the basic reproductive number $R_0$ is the largest eigenvalue of the next-generation matrix, then:

$$R_0=\frac{\alpha_1 S_0}{\beta+\tau_1}( 1+\frac{\alpha_2\beta}{\alpha_1(\tau_2+\gamma+\mu)})$$ 

\subsection{Parameter estimation}\label{estimation}

\paragraph*{Step 1:}
In this part, to estimate the initial conditions $As(t_0)$ and $I(t_0)$ and  parameters $\alpha_1$ and  $\alpha_2$ we  adapt the method developed by \cite{Liu2020}. 
 Let   $\gamma_1=\gamma+\mu$ and $CR(t)$ the cumulative number of reported  infectious cases at time $t$, defined by, 
\begin{equation*}
CR(t)=\int_{t_0}^t\tau_1 As(t)+\tau_2 I(t) dt
\end{equation*}

Let's assume that $CR(t)=\chi_1 \exp(\chi_2 t)-\chi_3$ with $\chi, \chi_2$ and $\chi_3$ three positive parameters.

By assuming that, $CR(t_0)=0 $  equation \eqref{eq:2} implies that:
\begin{equation}
 \exp(\chi_2 t_0)=\frac{\chi_3}{\chi_1} \mbox{ and then }t_0=\frac{1}{\chi_2}(\ln(\chi_3)-\ln(\chi_1)).
\end{equation}

Using equation \eqref{eq:2}, we have also:
\begin{align}
 CR'(t)&=\tau_1 As(t)+\tau_2 I(t)\\
 &=\chi_1\chi_2\exp(\chi_2 t)
\end{align}

Let's note $\tau=\frac{\tau_2}{\tau_1}$ and $H(t)=As(t)+\tau I(t)$. Then we have,  
\begin{equation}
\label{eq:H0}
H(t)=H(t_0)\exp(\chi_2 (t-t_0)),\mbox{with }
H(t_0)=\frac{\chi_3\chi_2}{\tau_1}. 
\end{equation}

In order to simplify the calculus, we will use the normalized  functions, $\frac{As}{H}$ and $\frac{I}{H}$. We have:
\begin{equation}
\label{eq:8}
\frac{As(t_0))}{H(t_0)}=1-\tau \frac{I(t_0)}{H(t_0)}. 
\end{equation}

Rewriting the third equation \eqref{eq:model}, with $H$ variable,

\begin{equation}
 \label{eq:6}
 \begin{array}{lcl}
 \displaystyle \frac{dI}{dt}&=&\beta H-(\tau_2+\beta \tau+\gamma_1)I
 \end{array}
\end{equation}

By assuming,  that $I(t)=I(t_0)\exp(\chi_2 (t-t_0))$ and  substituting in   equation \eqref{eq:6}, we obtain:

\begin{equation}
 \label{eq:7}
 \begin{array}{lcl}
 \displaystyle \chi_2I(t_0)&=&\beta H(t_0)-(\tau_2+\beta \tau+\gamma_1)I(t_0)
 \end{array}
\end{equation}

Equation \eqref{eq:7}, implies  
\begin{equation}
 \label{eq:I0}
\frac{I(t_0)}{H(t_0)}=\frac{\beta }{\chi_2+\tau_2+\beta \tau+\gamma_1}
 \end{equation}

By using equation \eqref{eq:8} and  \eqref{eq:I0}, we obtain: 
\begin{equation}
\label{eq:As0}
\frac{As(t_0))}{H(t_0)}=\frac{\chi_2+\tau_2+\gamma_1}{\chi_2+\tau_2+\beta \tau+\gamma_1}
\end{equation}

Let's assume that $\alpha_2=f\alpha_1$ with $f$ a fixed parameter bigger than $1$. The parameter  $\alpha_1$ is evaluated using $As(t)=As(t_0)\exp(\chi_2 (t-t_0))$ and the second  equation  of \eqref{eq:model} at $t_0$, we obtain:
\begin{align}
 \chi_2\frac{As(t_0)}{H(t_0)}&=\alpha_1 S_0(f\frac{I(t_0)}{H(t_0)}+\frac{As(t_0)}{H(t_0)})-(\beta +\tau_1)\frac{As(t_0)}{H(t_0)} \Leftrightarrow \\
 \alpha_1&=\frac{\chi_2+\beta+\tau_1}{(f\frac{I(t_0)}{As(t_0)}+1)S_0} 
 \end{align}

 and therefore using equations \eqref{eq:I0} and \eqref{eq:As0},
 
 \begin{align}
 \alpha_1&=\frac{(\chi_2+\beta+\tau_1)}{(\frac{f\beta}{\chi_2+\tau_2+\gamma_1}+1)S_0}
 \end{align}

\paragraph*{Step 2}
We hereby propose to estimate $\gamma$ and $\mu$. We notice that, $R(t)=\frac{\gamma}{\mu}D(t)$, for all  $t>0$. Let $\rho=\frac{\mu}{\gamma}$, $\rho$ is estimate using dead and recoveries  data.

Let $p$ the fraction of infectious (quarantined or not)  that become reported dead (\ie $1-p$ become reported recovered).
Thus $\rho=\frac{p\hat{\mu}}{(1-p)\hat{\gamma}}$, with $1/\hat{\mu}$ the average time to death and $1/\hat{\gamma}$ the average time to recover.
Therefore, 
\begin{equation}
p=\frac{\rho \hat{\gamma}}{\hat{\mu}+\rho \hat{\gamma}}. 
\end{equation}

\paragraph*{Step 3}
Parameters $\tau_1$ et $\tau_2$ was estimated using Metropolis-Hastings (MH) algorithm developed in the pymcmcst python package \cite{Miles2019}


\vspace{6pt} 

\end{document}